%% file: manuscript.tex
\documentclass[letter, 12pt]{article}
\usepackage{graphicx} 
\usepackage[T1]{fontenc}
\usepackage[english]{babel}
\usepackage{amsmath, amssymb}
\usepackage[table]{xcolor}
\usepackage{times}
\usepackage{helvet}
\usepackage{sectsty} 
\allsectionsfont{\sffamily} 
\usepackage[labelfont={bf,sf}]{caption} 
\usepackage{booktabs} 
\usepackage{hyperref} 
\usepackage{multirow}
\usepackage{pdflscape}
\usepackage[onehalfspacing]{setspace}
\usepackage{orcidlink} 
\usepackage{bbold}
\usepackage{enumitem}
\usepackage{natbib}
\usepackage{doi}
\usepackage{tikz}
\usetikzlibrary{positioning, arrows, fadings,decorations.pathmorphing,arrows.meta, shapes.arrows}
\tikzset{>=stealth'} 
\usepackage{longtable}
\definecolor{pigment}{rgb}{0.2, 0.2, 0.6}
\definecolor{lightgray}{gray}{0.9}
\hypersetup{
  colorlinks = true,    
  urlcolor   = pigment, 
  linkcolor  = black,   
  citecolor  = pigment  
}

\usepackage{geometry}
\newif\ifanonymize 
\newcommand{\anonymize}[1]{%
  \ifanonymize
    \phantom{#1}%
  \else
    #1%
  \fi
}
\anonymizefalse 

\title{\vspace{-3em}
\textsf{\textbf{
Handling Missingness, Failures, and Non-Convergence in Simulation Studies: A Review of Current Practices and Recommendations}
}}

\usepackage{authblk}

\author[1]{
\anonymize{Samuel~Pawel \orcidlink{0000-0003-2779-320X}}
}
\author[2,*]{
\anonymize{František~Bartoš \orcidlink{0000-0002-0018-5573}}
}
\author[3,*]{
\anonymize{Björn~S.~Siepe \orcidlink{0000-0002-9558-4648}}
}
\author[4,5,*]{
\anonymize{Anna~Lohmann \orcidlink{0000-0002-4004-4265}}
}
\affil[*]{
  \anonymize{
  Contributed equally
  }
}
\affil[1]{
  \anonymize{
  Epidemiology, Biostatistics and Prevention Institute,
  Center for Reproducible Science,
  University of Zurich
  }
}
\affil[2]{
  \anonymize{
  Department of Psychological Methods,
  University of Amsterdam 
  }
}
\affil[3]{
  \anonymize{
  Psychological Methods Lab, 
  Department of Psychology, 
  University of Marburg
  }
}
\affil[4]{
  \anonymize{
  Department of Clinical Epidemiology,
  Leiden University Medical Center
  }
}
\affil[5]{
  \anonymize{
  Department of Social Welfare,
  Ernst-Abbe University of Applied Sciences Jena
  }
}

\date{July 10, 2025} 

\begin{document}
\maketitle

\begin{abstract} 
\noindent Simulation studies are commonly used in methodological research for the empirical evaluation of data analysis methods. They generate artificial data sets under specified mechanisms and compare the performance of methods across conditions. However, simulation repetitions do not always produce valid outputs, e.g., due to non-convergence or other algorithmic failures. This phenomenon complicates the interpretation of results, especially when its occurrence differs between methods and conditions. Despite the potentially serious consequences of such ``missingness'', quantitative data on its prevalence and specific guidance on how to deal with it are currently limited. To this end, we reviewed 482 simulation studies published in various methodological journals and systematically assessed the prevalence and handling of missingness. We found that only 23\% (111/482) of the reviewed simulation studies mention missingness, with even fewer reporting frequency (92/482 = 19\%) or how it was handled (67/482 = 14\%). We propose a classification of missingness and possible solutions. We give various recommendations, most notably to always quantify and report missingness, even if none was observed, to align missingness handling with study goals, and to share code and data for reproduction and reanalysis. Using a case study on publication bias adjustment methods, we illustrate common pitfalls and solutions. 

\noindent \textsc{Keywords}: 
errors, meta-research, missing values, Monte Carlo experiments, reporting practices
\end{abstract}

\section{Introduction}
\label{sec:introduction}
A key goal of quantitative methodological research (e.g., statistics, psychometrics, bioinformatics, methods in ecology, econometrics, or machine learning) is to investigate the performance of data analysis methods. While formal analyses and mathematical proofs can clarify theoretical properties, they typically rely on assumptions that do not reflect real-world conditions. Consequently, methodologists often resort to \emph{simulation studies}, which allow empirical evaluation of methods under realistic scenarios.

In such studies, the methods under comparison are applied to artificial data sets that are simulated under a specified data-generating mechanism (DGM), followed by comparing their performance. As such, simulation studies are similar to controlled experiments because the underlying data-generating mechanism is usually known. Simulation studies are ubiquitous in methodological research. For example, \citet{Morris2019} found that 199/264 = 75\% of all articles published in volume 34 of the journal \emph{Statistics in Medicine} contained at least one simulation study. In our literature review, we found that in volume 118 of \emph{Journal of the American Statistical Association} 186/200 = 93\% (!) of articles reported at least one simulation study. 

Methodological rigor in simulation studies is crucial because they often guide data analysis and statistical method selection. Consequently, simulation studies inform scientific, medical, and policy decisions, potentially for decades to come. Several guidelines and tutorials have been published to support high-quality simulation studies \citep[e.g.,][]{Hoaglin1975, Hauck1984, Burton2006, Sigal2016, Morris2019, Chalmers2020, Boulesteix2020B, Chipman2022, Kelter2024, Siepe2024, Williams2024}. 

A key issue lacking detailed guidance is when simulation repetitions fail to produce valid outputs (e.g., parameter estimates, standard errors, confidence intervals, predictions, \textit{p}-values, sample sizes, performance measure estimates, valid data sets) required to assess method performance. For example, this could happen if an optimization algorithm for estimating model parameters does not converge and therefore does not produce parameter estimates. Throughout this paper, we use \emph{missingness} as an umbrella term for non-convergence, improper solutions, ill-defined data sets, ill-defined performance metrics, run time errors, and similar problems. The terminology used in the literature to describe this phenomenon is ambiguous. \citet{White2010}, \citet{Hennig2018}, \citet{Morris2019}, \citet{Gasparini2021}, \citet{Pawel2024}, and \citet{Siepe2024} use the term ``missing value'' to describe these problems, while \citet{Wuensch2024} note that this conflicts with the definition of missing values from the missing data literature and prefer the term ``failure'', which was also used by \citet{Burton2006}. However, ``failure'' could be interpreted as an error in the simulation study (e.g., DGM or method implementation) itself, which is often not the cause of missingness. We hence use ``missingness'' for its neutrality, colloquial suitability, and consistency with prior literature.

Missingness in simulation studies is likely to become more prevalent with the increasing complexity of statistical, machine learning, and artificial intelligence methods, and the feasibility of large-scale simulations. Missingness has recently been identified as a barrier to replicability \citep{Luijken2024}, and a potential source of researcher's degrees of freedom that enable questionable research practices \citep{Pawel2024}. Limited guidance on handling missingness also affects method benchmarking with real data sets \citep{Boulesteix2013, Boulesteix2017, Boulesteix2020, Niessl2021, Wuensch2024}. Although we focus on simulation studies, many of the problems and solutions are directly transferable.

It has been generally recommended to report frequency and patterns of missingness \citep{Burton2006, Morris2019, Chalmers2020, Giordano2020, Kelter2024, Wuensch2024} as well as to pre-specify how these cases will be handled \citep{Kuribayashi2014,Pawel2024, Siepe2024, Luijken2024, Williams2024}. However, apart from reporting and pre-specification, it is not clear how exactly missingness should be handled when it occurs. Simply omitting missing observations from the analysis can distort the conclusions of the study. For instance, if a method fails to converge on the most challenging data sets, excluding results only for the non-convergent method while keeping the results for other methods may bias the results in favor of the excluded method. Rather than omitting missing observations, one could alternatively simulate new data sets until all methods converge, adjust the parameters of the method until it converges, or impute missing values using the worst-case or mean performance -- each of these approaches could impact the assessment of method performance differently. So far, however, the problem has not been thoroughly investigated, and researchers have instead mostly come up with ad hoc solutions on a case-by-case basis.

A striking example of how missingness can distort conclusions is the simulation study that advocated the ``ten events per variable'' rule for determining the sample size in logistic regression \citep{Peduzzi1996}. Cited over 8'000 times, this rule is an influential tool in evaluating medical studies. However, later research showed that the study's findings were driven by how non-convergent iterations were handled -- specifically those affected by ``complete separation'', which arises more often in conditions with few events per variable \citep{Steyerberg2011, vanSmeden2016, vanSmeden2018}. This example demonstrates how suboptimal handling of missingness can have far-reaching negative consequences in the practical implementation of statistical methods.

The prevalence of missingness in simulation studies -- and the extent of the problem in the literature -- remains unclear. In their review of articles from \emph{Statistics in Medicine}, \citet{Morris2019} found that 14\% (12/85) of simulation studies reported convergence as a performance measure when applicable, similarly \citet{Siepe2024} found that 19\% (19/100) of reviewed articles from methodological journals in psychology reported convergence. \citet{Hinds2018} reviewed simulation studies on methods for longitudinal patient-reported outcome data and found that more than half of the surveyed studies did not mention non-convergence. However, beyond these studies, to our knowledge, no comprehensive quantification of missingness prevalence or handling strategies exists.

This paper has two primary goals: First, we aim to systematically assess how the issue of missingness in simulation studies is currently reported and handled in the methodological research literature. Second, we aim to provide practical recommendations for methodological researchers on approaching missingness. To this end, we systematically reviewed research articles published in various methodological journals, summarized in Section~\ref{sec:literature-review}. We then propose a classification of different types of missingness  (Section~\ref{sec:classification}) and outline approaches to dealing with them, including their strengths and limitations (Section~\ref{sec:approaches}). In Section~\ref{sec:case-studies}, we use a case study on methods for publication bias adjustment in meta-analysis to illustrate how missingness can affect results and suggest ways of handling it. Finally, we provide practical recommendations (Section~\ref{sec:recommendations}) and conclude with a discussion of our findings, limitations, and implications (Section~\ref{sec:discussion}).

\section{Literature review}
\label{sec:literature-review}
We systematically reviewed recent issues of the \emph{Journal of the American Statistical Association} (JASA), \emph{Research Synthesis Methods} (RSM), \emph{Statistics in Medicine} (SiM), and \emph{Psychological Methods} (PM). These journals were selected for their prominence in quantitative methodological research across various fields, and because they align with the authors' statistical expertise. 

\subsection{Sampling and coding procedure}
The reviewed issues reported 800 research articles, out of which 482 (60\%) contained at least one simulation study. Reporting details were extracted by going through journal issues in reverse chronological order, starting with the last issue of 2023. Articles were included until each coder (all four authors) had coded at least one issue per journal and a minimum of 100 studies per journal were coded that fulfilled the inclusion criteria stated below, or until the first issue of 2018 was reached.%
\footnote{For RSM, the first issue of 2018 was reached with 94 simulation studies. For PM and SiM, one issue was skipped due to our coder allocation approach. Due to a minor miscount, 98 instead of 100 simulation studies were coded for PM.}
Only original research articles, reviews, and tutorials were considered, and editorials, commentaries, book reviews, corrections, or letters-to-editors were excluded. The preregistered literature review protocol with further details on the extraction procedure can be accessed at \anonymize{\url{https://doi.org/10.17605/OSF.IO/PMV2J}}. Additional numerical and visual summaries of our results are available at \anonymize{\url{https://github.com/SamCH93/missSim}}. The coding scheme was tested on eight pilot studies that each coder coded, these were excluded from the final results. Each coder rated their overall confidence in coding each study as ``low'', ``medium'', or ``high''. To assess inter-rater agreement, 50 randomly selected studies with ``low'' or ``medium'' confidence were coded by a second coder. We present the results of the inter-rater agreement in the \hyperref[sec:appendix]{Appendix}. Overall, we found inter-rater agreement to be satisfactory, with a mean of 80.6\% for the most challenging simulation studies. This likely represents a lower bound, as the majority of studies were easier to code. In case of disagreement, the first coder's coding was used for all analyses reported below.

\begin{figure}[!htb]
    \centering
    \includegraphics[width=1\linewidth]{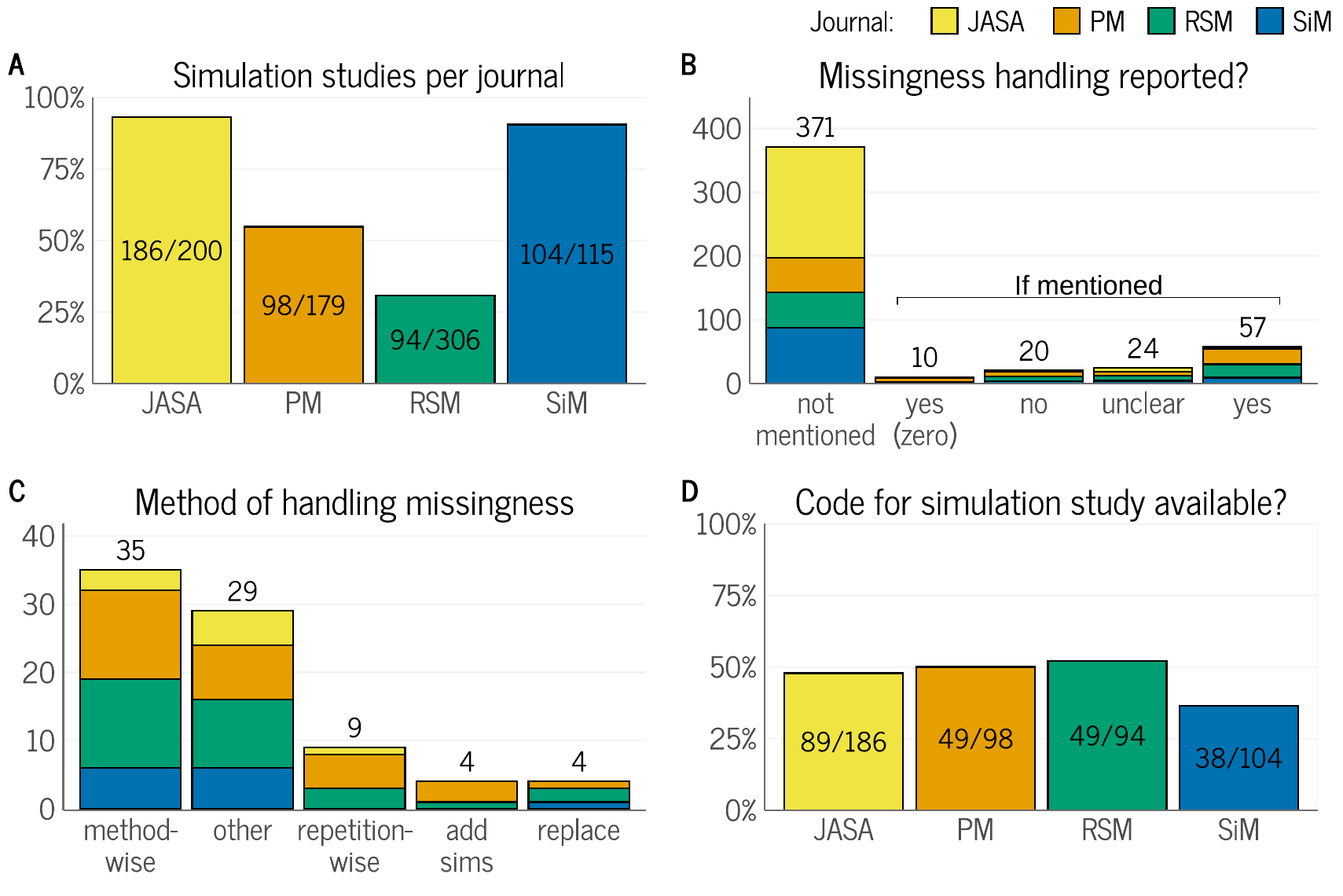}
    \caption{Main results of the literature review of the \emph{Journal of the American Statistical Association} (JASA), \emph{Psychological Methods} (PM), \emph{Research Synthesis Methods} (RSM), \emph{Statistics in Medicine} (SiM). In panels A and D, we show per-journal percentages on the y-axis and proportions of articles with simulation studies and simulation studies with code, respectively, as fractions inside the bars. Panel B shows how many articles mentioned missingness and, if so, whether they described how it was handled. ``Yes (zero)'' refers to articles that explicitly reported that no missingness occurred. In panel C, ``method-wise'' and ``repetition-wise'' refer to the deletion of missing values, ``add sims'' refers to conducting additional simulation repetitions until a desired number is achieved, ``replace'' refers to the replacement of a method with another one. For panel D, we checked whether code for the simulation study and not just the implementation of the method was available. We did not check whether we could run the code.}
    \label{fig:lit-review}
\end{figure}

\subsection{General findings}
Figure~\ref{fig:lit-review} shows the main results of our literature review. 
Simulation studies were most frequent in JASA (186/200 = 93\%), followed by SiM (104/115 = 90\%), PM (98/179 = 55\%), and RSM (94/306 = 31\%). These proportions illustrate the ubiquity of simulation studies in methodological research. Of the 482 simulation studies, 111 (23\%) mentioned missingness (see panel B). This is consistent with prior reviews:  \citet{Morris2019} and \citet{Siepe2024} found that 14\% and 19\% of studies, respectively, reported convergence as a performance metric (which is one way of mentioning missingness).

\subsection{Acknowledgement of missingness}
Studies mentioning missingness typically did so in the results section in text form. The most common ways of summarizing missingness were by reporting (absolute or relative) frequency per method and condition (50/111 = 45\%), or by merely acknowledging it without providing explicit quantification (19/111 = 17\%). For example, \citet{hoyer2018} provided a table with the number of converged repetitions per condition and method. 

\subsection{Handling missingness}
Only around half of the studies that acknowledged the existence of missingness also elaborated on how missingness was handled (57/111 = 51\%), with some additional studies explicitly reporting that no/zero missingness occurred (10/111 = 9\%). In the remaining studies, authors either did not specify the handling approach (20/111 = 18\%) or it was unclear (24/111 = 22\%, see Panel B).

For the studies that elaborated on how missingness was handled when it occurred, we recorded the typical strategies of deletion (``method-wise'' or ``repetition-wise''), ``additional simulation repetitions'', and ``method replacement'' (see panel C of Figure~\ref{fig:lit-review}), each will be discussed in more detail in Section~\ref{sec:approaches}. This includes some studies in which the wording was not fully clear, but some handling was reported clearly (overall $n$ = 81). Hence, these studies include studies coded as ``unclear'' in panel B. 
Method-wise deletion denotes that only those per-method cases containing missing values were omitted in the performance evaluation. This was the most commonly applied strategy among studies that provided any details on missingness handling (used in 35/81 = 43\% of the studies). Method-wise deletion was furthermore the most common strategy that was not explicitly reported but implicitly assumed by the coders (unquantified). Repetition-wise deletion (9/81 = 11\%) involves not only deleting the missing value itself, but also values from the other methods in the same simulation repetition.
Only a small number of studies chose to perform additional simulation repetitions to compensate for the loss of deleted repetitions (4/81 = 5\%) or replaced the missing values (4/81 = 5\%), for example with the output of a different method. 
Other strategies (used in 29/81 = 36\% of the studies) often combined two handling approaches. For example, \citet{Liu2022} either performed additional simulation repetitions or did not interpret the performance of a method in case of a very high non-convergence rate.

\subsection{Justifications for missingness handling}
As guidance for handling missingness in simulation studies is limited, we were interested in the justification that the authors provided. In most cases, no clear justification was provided for the choice of missingness handling, or the rationale was unclear (65/81 = 80\%). When justifications were given, they were typically based on reasoning (14/20 = 70\%) rather than, e.g., referring to another article.%
\footnote{These 20 studies that provided some justification included four studies for which the justification was (partially) unclear.}
An example way of reasoning can be found in \citet{Leahy2018}. In an effort to ``eliminate any potential bias due to differing simulations'' (p.~446), these authors excluded all values of all models from simulation repetition where at least one model failed to converge. Another example encompassing multiple positive aspects of reporting is \citet{Pustejovsky2019}. They provided different causes of non-convergence in their simulations and explained their reasons for choosing how to deal with them. Additionally, they created an extensive supplement including per-condition non-convergence rates for a method with relatively high non-convergence rates as well as code to reproduce the simulation study. 

\subsection{Code sharing}
Code sharing facilitates reproducibility and transparency \citep{Chalmers2020, Siepe2024, Luijken2024}, particularly in understanding how missingness was handled. We therefore tracked the availability of software code for reproducing the simulation study (see panel D). We did not assess the code regarding missingness handling, as this was infeasible due to the broad scope of the review. Overall, simulation code was available for less than half of all simulation studies (225/482 = 47\%). This result is similar to the findings of previous literature reviews in psychology and biostatistics \citep{Morris2019, Kucharsky2020, Siepe2024}. While code availability was around 50\% for most reviewed journals, SiM stood out with a substantially lower code availability (37\%). 

A positive example is \citet{Weber2021}, who documented simulation errors and their handling in the supplement. They also shared code and output files enabling reproduction of their results and all warning messages. 

\section{A classification of missingness}
\label{sec:classification}
To understand missingness and explore the potential impact of approaches to dealing with it, it is useful to classify missingness into different types. We base our classification on the steps of a simulation study where missingness can occur, see Figure~\ref{fig:schema} for a schematic illustration and Table~\ref{tab:classification} for a summary.

\begingroup
\renewcommand{\arraystretch}{1.3} 
\begin{figure}[!htb]
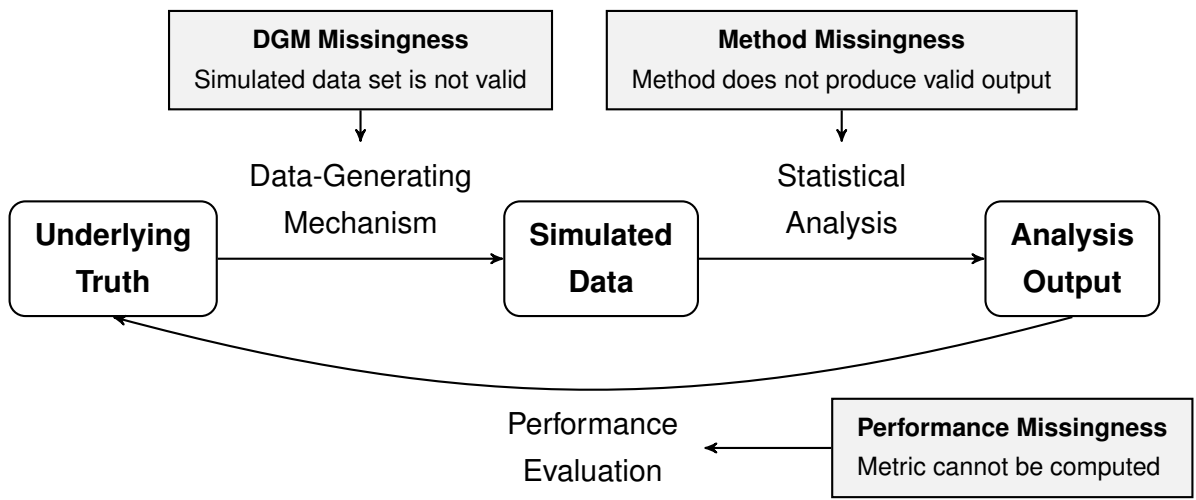

\centering
{\sf 
\include{missingness-classification-tikz}
}
\caption{Schematic illustration of types of missingness in simulation studies.}
\label{fig:schema}
\end{figure}
\endgroup


\begin{table}[!htb]
    \centering
    \caption{Classification of types of missingness in simulation studies.}
    \label{tab:classification}
    {\small
    \begingroup
    \renewcommand{\arraystretch}{1.5} 
    \begin{tabular}{p{0.25\linewidth} | p{0.25\linewidth} | p{0.4\linewidth}}
    \toprule
     \multicolumn{1}{c}{\textbf{Missingness type}} & \multicolumn{1}{c}{\textbf{Implications}} & \multicolumn{1}{c}{\textbf{Examples}} \\
     \midrule
     \rowcolor{lightgray}
     \textbf{DGM missingness} \newline
     DGM produces an invalid data set & 
     ~ \newline 
     Simulated data set is not usable for analysis and performance evaluation
     &
     \vspace*{-0.2cm}
     \begin{itemize}[topsep=-0.5cm, leftmargin=0.3cm]
     \item DGM produces non-positive semidefinite covariance matrices 
     \item DGM produces binary outcomes with only zeros or only ones
     \end{itemize}
     \\
     
     \textbf{Method missingness} \newline
     Method produces invalid output from valid data set &
     ~ \newline 
     Missing output has to be excluded or replaced to estimate method performance
     &
     \vspace*{-0.2cm}
     \begin{itemize}[topsep=-0.5cm, leftmargin=0.3cm]
     \item Iterative method does not converge, converges to local minima or unrealistic values (e.g., mean outside of data range)
     \item Method produces inadmissible outputs (e.g., negative variances, absolute correlations > 1, parameter estimate of infinity)
     \item Multistage method produces boundary case (e.g., variable selection method selects no variable)
     \item Method encounters computational limits (e.g., memory overflow, time-out)
     \end{itemize} 
     \\
     
     \rowcolor{lightgray}
     \textbf{Performance missingness} \newline
     Performance metric cannot be estimated from valid method output & 
     ~ \newline 
     Performance cannot be assessed for missing metric
     &
     \vspace*{-0.2cm}
     \begin{itemize}[topsep=-0.5cm, leftmargin=0.3cm]
     \item  Predicted probabilities of 0/1 lead to log score of infinity
     \item  Calibration slope cannot be estimated due to constant predictions
     \item  Mean squared error numerically explodes due to extreme predictions 
     \end{itemize} 
     \\ 
     \bottomrule
    \end{tabular}
    \endgroup
    }
\end{table}

\subsection{DGM missingness}
A repetition of the specified data-generation mechanism (DGM) may fail to produce a data set, or may produce a data set that is in some way ``ill-defined''. For instance, \citet{Johal2023} used a simulation procedure for generating polychoric correlation matrices, which occasionally produced non-positive semidefinite matrices -- invalid as correlation matrices. As a result, those data sets could not be analyzed or used to assess performance. To address this, \citet{Johal2023} repeated sampling until a valid matrix was generated and reported the frequency of such failures.

DGM missingness is typically less problematic, as it rarely biases performance estimates. However, transparent reporting remains essential. Otherwise, it may be difficult to assess whether the (possibly modified) DGM remains relevant in practice (e.g., whether it changed the true estimand). It may also hinder replication or cross-study comparison.

\subsection{Method missingness}
When a method is applied to a simulated data set, the method may fail to produce a valid output (e.g., parameter estimate, confidence interval, prediction, \textit{p}-value, sample size) needed to estimate its performance. This may be due to an algorithmic failure in the fitting of the method, such as the non-convergence of an algorithm. In this case, the performance of the method can usually not be estimated, unless the missing output is replaced or excluded. For example, \citet{vanZundert2020} encountered errors and non-convergence with one method. They addressed this by tweaking the optimization parameters of the method and excluding the 1.5\% cases that still did not converge. 

\citet{Wuensch2024} discuss technical reasons for why method missingness can occur in method comparison studies; a method may (1) use up all the available memory and cause a system crash, (2) not complete computations within a given time limit, (3) not be able to perform required computations (not converge, give ill-defined output, etc.).  The first two can be resolved by increasing computational resources, implying a ``true'' but unobserved value. The third lacks a well-defined true value. This distinction determines whether approaches to missing value handling from ordinary data analysis, such as multiple imputation, make conceptual sense, which will be discussed further in Section~\ref{sec:imputation}.

In some cases, the distinction drawn by \citet{Wuensch2024} may lack clarity. For example, it may be unclear whether a method could in principle converge, but fails to do so within the specified number of iterations and convergence criterion, or whether it is inherently unable to converge (e.g., because the objective function being optimized is unbounded). Additionally, the distinction depends on whether a ``method'' is defined as a general approach (e.g., maximum likelihood logistic regression) or a specific implementation (e.g., the \texttt{glm} implementation of R with default arguments). For a given data set, there may be a well-defined logistic regression maximum likelihood estimate, but the \texttt{glm} function may still fail to converge.
Similarly, in the simulation study of \citet{Ross2023}, different implementations of the same method in either R or SAS had different non-convergence rates. Depending on whether one considers these implementations to be the same or different methods, they may or may not be able to converge for certain data sets. 

There are also cases where the distinction between DGM and method missingness is ambiguous. For example, in a simulation study evaluating the performance of logistic regression, a simulated data set may be completely separated (i.e., the outcome can be perfectly predicted from the covariates), in which case ordinary maximum likelihood logistic regression cannot converge because the maximum likelihood estimate does not exist. Depending on the goals of the simulation study, this may be defined as DGM missingness (e.g., when only data sets without complete separation are of interest) or method missingness (e.g., when methods that can handle complete separation are also included in the simulation study).

Despite these ambiguities, it is clear that method missingness can pose a major challenge for interpreting simulation results, particularly when missingness differs between the methods being compared. The missingness handling approaches described in the Section~\ref{sec:approaches} will therefore mostly focus on method missingness, and when we use the term ``missingness'' without preceding DGM / method / performance, we refer to this missingness type. 

\subsection{Performance missingness}
A performance metric may be incomputable or undefined for otherwise valid method outputs. For instance, if a prediction method predicts a binary outcome with absolute certainty (i.e., predicts a probability of 0 or 1) and the observed value is the opposite of the predicted outcome, the log score (negative log likelihood) becomes infinity. Another example is the simulation study by \citet[not part of our review]{Dunias2024}, which compared different methods for hyperparameter tuning of prediction models. For certain data sets, the LASSO method selected no predictors, producing constant predictions that precluded estimating performance via the calibration slope \citep{VanCalster2019}.
Consequently, the performance of the method cannot be assessed with respect to the missing performance metric. To conservatively estimate method performance, \citet{Dunias2024} replaced the missing calibration slopes by the maximum calibration slope of the corresponding condition.

We suspect that performance missingness often goes unreported, as researchers may simply switch metrics. The distinction between method and performance missingness can also sometimes be ambiguous. For example, predicted probabilities near 0 or 1 may be represented as exact 0 or 1 due to limited numerical precision. Consequently, the log likelihood of an incorrect prediction becomes infinity. Depending on the definition, this scenario could be viewed as either method missingness or performance missingness. The study by \cite{zhang2023} comparing various methods for Poisson prediction is an example where this happened. The results table indicating average root mean squared error of prediction (RMSE) across methods and conditions frequently contains ``Inf(Inf)'' especially in more challenging cases, presumably because the RMSE ``exploded'' due to extreme predictions.

\section{Approaches for assessing and handling missingness}
\label{sec:approaches}
\begin{figure}[!htb]
    \centering
    \includegraphics[width=0.95\linewidth]{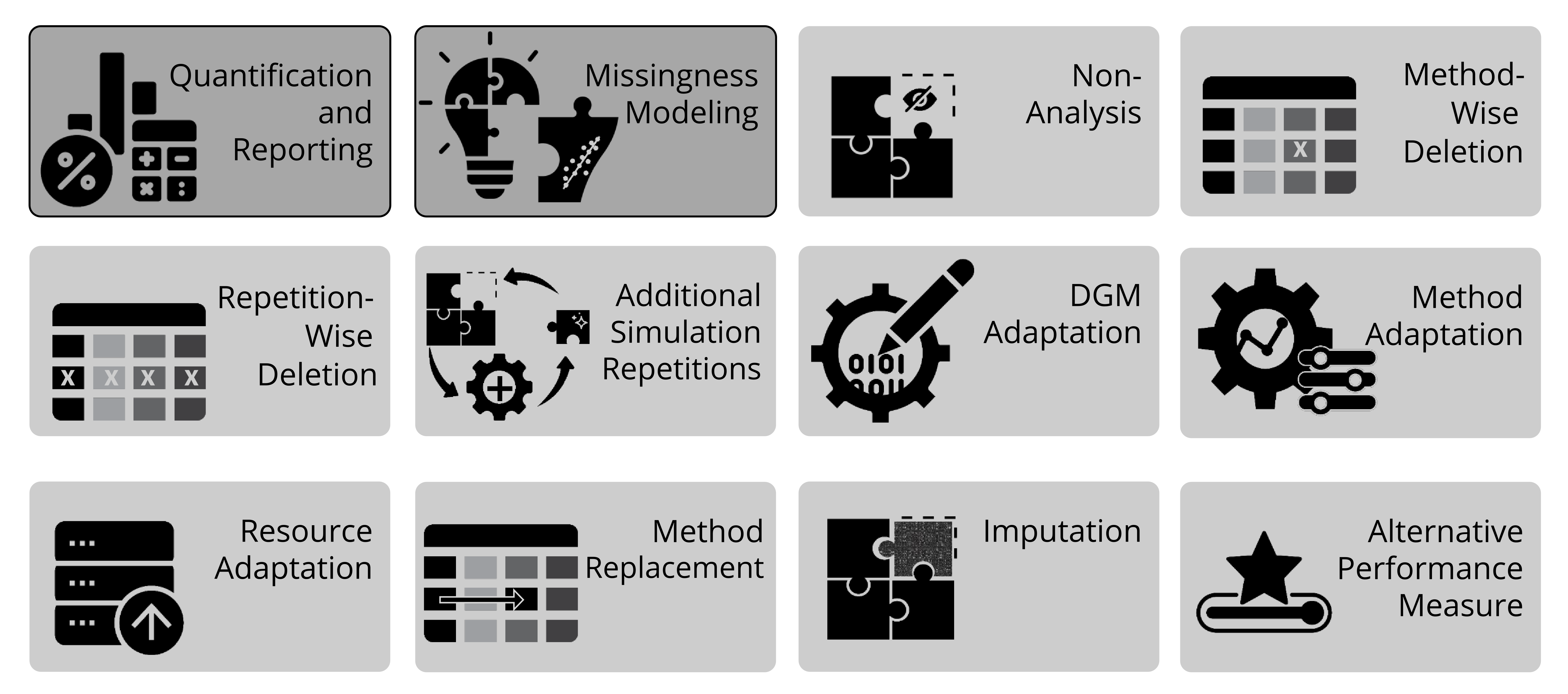}
    \caption{Symbolic illustration of approaches for assessing missingness (with black borders) and handling missingness (without black borders) in simulation studies.}
    \label{fig:handling}
\end{figure}

We now outline approaches for assessing and handling missingness in simulation studies along with examples from our review. Note that these examples serve only as illustrations and are not necessarily intended as recommendations for how missingness should be approached. A visual summary of the different approaches is given in Figure~\ref{fig:handling}, and a more detailed summary is given in Table~\ref{tab:approaches}. 

\newpage

{\small
\begin{longtable}{p{0.35\linewidth} p{0.6\linewidth}}
    \caption{Summary of approaches for assessing and handling missingness in simulation studies.} \label{tab:approaches} \\
    \toprule
    \rowcolor{white}
    \multicolumn{1}{c}{\textbf{Approach}} & \multicolumn{1}{c}{\textbf{Advantages and disadvantages}} \\
    \hline
    \endfirsthead
    \toprule
    \rowcolor{white}
    \multicolumn{1}{c}{\textbf{Approach}} & \multicolumn{1}{c}{\textbf{Advantages and disadvantages}} \\
    \hline
    \endhead
    \bottomrule\endfoot

    \multicolumn{2}{c}{\textit{Assessing missingness}} \\
    \midrule

    \rowcolor{lightgray}
    \textbf{Quantification and reporting} \newline
    Report frequency of missingness (overall / by method / by method-condition / maximum overall or per method or per condition), visualize missingness rates to detect multivariate patterns & 
    \begin{itemize}[nosep]
        \item[+] Indicates whether missingness is potentially an issue, allows the reader to factor in this information when interpreting the results
        \item[--] Does not provide a way to analyze performance when missingness occurs
    \end{itemize} \\
    
    \textbf{Missingness modeling} \newline
    Fit model to understand occurrence of missingness (e.g., linear/logistic regression or decision tree)& 
    \begin{itemize}[nosep]
        \item[+] Can lead to a better understanding of missingness mechanism
        \item[--] Does not provide a way to analyze performance when missingness occurs, interpretation might be limited due to the inadequacy of the model for capturing the missingness mechanism 
    \end{itemize} \\

    \multicolumn{2}{c}{\textit{Handling missingness}} \\
    \midrule

    \rowcolor{lightgray}
    \textbf{Non-analysis} \newline
    Do not analyze/interpret performance of conditions and/or methods with too much missingness & 
    \begin{itemize}[nosep]
        \item[+] Resembles real-world practice in that a method is not even considered for certain conditions, no misinterpretation based on analysis of non-missing values
        \item[--] Performance for certain methods/conditions is completely missing, poor performance of a method can be concealed if it is connected to missingness
    \end{itemize} \\

    \textbf{Method-wise deletion} \newline
    Omit missing value when a method shows missingness but keep non-missing values of the other methods in the same repetition &
    \begin{itemize}[nosep]
        \item[+] Simple, avoids deletion of non-missing methods
        \item[--] Performance estimates 
        might be non-representative of the intended DGM, different methods' performance estimates can be based on different data, the resulting method comparison might be difficult to interpret, different methods may have an unequal number of repetitions
    \end{itemize} \\

    \rowcolor{lightgray}
    \textbf{Repetition-wise deletion} \newline
    Omit missing value when a method shows missingness and also omit non-missing values of the other methods in the same repetition &
    \begin{itemize}[nosep]
        \item[+] Simple, different methods have the same number of repetitions, method comparison based on the same data
        \item[--] Performance estimates 
        might be non-representative of the intended DGM, artificially increases uncertainty of performance estimates of the non-missing methods 
    \end{itemize} \\
    
    \textbf{Additional simulation repetitions} \newline
    Perform additional simulation repetitions to compensate for omitted repetitions, can be combined with method-wise or repetition-wise deletion or used to address DGM missingness &
    \begin{itemize}[nosep]
        \item[+] Different methods have the same desired number of repetitions, comparison based on the same data (when combined with repetition-wise deletion)
        \item[--] Performance estimates 
        might be non-representative of the intended DGM, computationally more intensive, hard to achieve with high missingness rates
    \end{itemize} \\

    \rowcolor{lightgray}
    \textbf{DGM modification} \newline
    Change the DGM so that no missingness occurs &
    \begin{itemize}[nosep]
        \item[+] Removes missingness
        \item[--] Relevant DGMs may be excluded, often requires re-running simulation study and ad hoc decisions on new DGM, could require multiple looks at results 
    \end{itemize} \\

    \textbf{Method adaptation} \newline
    Modify methods such that missingness no longer occurs (e.g., modify tuning parameters, number of iterations, convergence criterion, starting values), either only in case of missingness or for all simulations & 
    \begin{itemize}[nosep]
        \item[+] Aligns closer to how in practice data analysts would handle missingness
        \item[--] Harder to interpret, not always possible, could require multiple looks at results, potentially not impartial to methods that are not modified
    \end{itemize} \\

    \rowcolor{lightgray}
    \textbf{Resource adaptation} \newline
    In case of memory overflow or time-outs, increase computational resource allocation (e.g., CPU, run time, memory) & 
    \begin{itemize}[nosep]
        \item[+] Aligns closer to practice where data analysts have more resources to implement a single repetition of a single method
        \item[--] Expensive regarding time and computational resources, not always possible
    \end{itemize} \\

    \textbf{Method replacement} \newline
    Replace missing values with values from an alternative method (e.g., current gold standard, baseline method or more robust method variant) &
    \begin{itemize}[nosep]
        \item[+] Missing values are replaced, no additional simulation repetitions required, can emulate what would be done in practice
        \item[--] Harder to interpret, requires choice of an alternative method, introduces a new source of uncertainty, does not evaluate ``pure'' method performance
    \end{itemize} \\

    \rowcolor{lightgray}
    \textbf{Imputation} \newline
    Impute missing values (e.g., by worst-case, conditional mean, or other summary) &
    \begin{itemize}[nosep]
        \item[+] Missing values are replaced, no additional simulations required
        \item[--] Harder to interpret, requires choice of imputation method, classical imputation methods are not appropriate if there is no ``true'' underlying value that is missing
    \end{itemize} \\

    \textbf{Alternative performance measure} \newline
    In case of performance missingness, switch to an alternative performance measure that does not show missingness (e.g., trimmed mean, mean absolute deviation, ranks, winsorization, etc.) &
    \begin{itemize}[nosep]
        \item[+] Method performance can be assessed
        \item[--] Relevant performance measures may be excluded, may lead to selective reporting and outcome switching, may obscure issues with methods
    \end{itemize} \\
\end{longtable}
}

\subsection{Assessing missingness}

\subsubsection{Quantifying and reporting missingness}
Quantifying the prevalence of missingness is a basic but crucial step in the analysis and reporting of simulation studies \citep{Burton2006, Morris2019, Chalmers2020, Giordano2020, Siepe2024, Pawel2024, Kelter2024, Wuensch2024}. It is important to note that detecting missingness may be nontrivial, for instance, when an algorithm converges to a nonsensical solution. We refer to \citet{White2024} and \citet[Section 4.2]{Morris2019} for recommendations on how to check simulation study results for missingness. In the following, we will assume that missingness can be reliably detected.

Even if no missingness occurs, it is advisable to report this explicitly. Otherwise, readers cannot differentiate between the absence of missingness and the failure to report (or even check). If missingness occurs, researchers can quantify it in several ways, the most fine-grained being to report the proportion of missingness by method and condition. This allows readers to put the results into context and reflect on whether missingness may lead to bias or just more imprecise results. Frequencies may be reported in a table or visualized in a graph, the latter can often help detect missingness patterns \citep{Gasparini2021, Templ2023, Tierney2023}. While reporting missingness by method and condition is the most informative, such detailed summaries may have to be moved to an appendix or online supplement if space is limited. Reporting missingness overall, per method, per condition, or a maximum proportion of missingness per method can serve as alternatives for the main text. 

A good example is \cite{castro-alvarez2022}, who differentiated between (i) errors and warnings, (ii) non-convergence, and (iii) resource limitations. They quantified the frequency of each per method and estimation approach in a table. Additionally, they visualized the number of successful analyses per condition.
\cite{huang2021} used a more minimal approach by adding a column to results tables showing the number of converged cases -- a simple practice that could become standard.
Another example highlighting the importance of missingness quantification per method and condition is \cite{hoyer2020}. Individual conditions for one method had as low as 0.3\% converged repetitions, casting doubt on its practical utility.

\subsubsection{Missingness modeling}
If there are many simulation conditions, it can also be helpful to fit a ``meta-model'' with missingness as the outcome and the methods and simulation factors as covariates \citep[see e.g.,][for meta-models in simulation studies]{Skrondal2000, Chipman2022}. For example, linear/logistic regression or decision trees \citep{Tierney2015} may be used for this purpose. Meta-modeling can help identify methods and conditions for which missingness is more likely to occur, providing insights into potential solutions. At the same time, inferences drawn from these models are dependent on the adequacy of the model assumptions and may be misleading if a meta-model model is inadequate \citep{morris_meta-models_2024}. 
Meta-models may also overcomplicate cases where missingness is concentrated within a few specific simulation settings, in which case descriptive statistics and graphs are sufficient. Moreover, there may be technical difficulties, for example, a logistic regression meta-model may itself fail to converge due to complete separation (because a method always/never converges). Likely for these reasons, we found no examples of missingness meta-models in our review.

\subsection{Handling missingness}
While reporting or modeling missingness allows for a better understanding and interpretation of the simulation results, it does not provide a solution to how they should be analyzed. 
We now describe approaches for handling missingness. Unlike missingness assessment approaches which can be combined, only one handling strategy is typically used to analyze method performance. Of course, sensitivity analyses can be used to compare alternatives (see Section~\ref{sec:recommendations}).

\subsubsection{Non-analysis of conditions or methods with missingness}
Researchers may opt not to analyze performance further if a method or condition shows too much missingness. For example, \citet{Molenaar2021} chose not to report and interpret the results for conditions with $<$ 15\% convergence for some methods. This approach avoids being misled by missingness but is also unsatisfactory as not much can be learned about method performance \citep{Morris2019}. Furthermore, there is a risk that poor performance of a method is concealed if it is connected to missingness.

\subsubsection{Method-wise and repetition-wise deletion of missing values}
A simple strategy to estimate method performance in the presence of missingness is omitting the missing values via method-wise or repetition-wise deletion. However, both may implicitly lead to a different DGM than originally specified, since there is conditioning on non-missingness. This implicit DGM may not be representative of the specified DGM, especially if missingness is dependent on the characteristics of the simulated data sets (which it often is). To investigate this, it may be useful to look at descriptive statistics of the simulated data sets conditional on deletion. For example, if in a DGM a sample size is first simulated based on which a data set is then simulated, it may be useful to assess whether, within a condition, the average simulated sample size after deletion matches the expected sample size of the specified sample size distribution.

An additional problem with method-wise deletion is that if missingness varies between methods, their estimated performance is based on different underlying data. This can make it difficult to compare the precision of performance estimates (e.g. with Monte Carlo standard errors) as a different number of repetitions may be available for each method. It can further distort the results. For example, if a particular method fails to converge on the most challenging data sets (e.g., data sets with small sample sizes), method-wise deletion may result in an over-optimistic assessment of its performance \citep{Siepe2024, Wuensch2024}. This is usually less of an issue with repetition-wise deletion since performance estimates are based on the same data for all methods, yet the implicit DGM underlying these performance estimates may be quite different from the specified DGM as there is a conditioning on simultaneous non-missingness of all methods. It is also essential to recognize that different missingness rates across methods are relevant for their practical utility. Finally, repetition-wise deletion also discards actual non-missing observations, thereby artificially increasing Monte Carlo uncertainty. 

An interesting example from our literature review is the study by \citet{Seo2023} on multiple imputation in prediction modeling. These authors discuss the trade-off between method-wise and repetition-wise deletion, noting that repetition-wise deletion enables a fairer method comparison while method-wise deletion leads to much fewer omitted repetitions in conditions where multiple imputation does not converge. In the end, they decided on a combination of the two approaches, using repetition-wise deletion when the convergence rate of multiple imputation was below 20\% in a condition, and method-wise deletion if it was above 20\%.


\subsubsection{Additional simulation repetitions}
To avoid an increase in Monte Carlo uncertainty from omitting simulation repetitions with missing values, additional simulations can be run until the desired number of complete repetitions is reached \citep{Paxton2001, Burton2006, Chalmers2020}. This can be computationally intensive and difficult to implement when missingness rates are high. In case of method missingness, the approach is typically combined with repetition-wise deletion. However, in principle, it could also be combined with method-wise deletion to avoid omitting non-missing values. That is, one could still retain repetitions where some methods are missing and perform additional simulations until all methods have at least the desired number of non-missing values. This may result in different numbers of non-missing values among methods, as some methods may have more non-missing values than the desired number of repetitions. Finally, when used to address DGM missingness, the additional simulation approach resembles repetition-wise deletion as no performance is estimated for all methods for invalid data sets.

To give an example from our literature review, \citet{Lai2022} used a repetition-wise deletion approach and performed additional simulation repetitions until they reached 5'000 repetitions where all compared methods converged to compensate for the omitted repetitions. However, this approach does not address the change in the implicit DGM, and could therefore still distort the results. 

\subsubsection{DGM adaptation}
Another option is to modify the DGM so that missingness no longer occurs \citep{Boomsma2013}. For example, if a method struggles to converge in conditions with small sample sizes, these conditions could be removed and replaced with conditions using larger sample sizes. While this approach can be helpful in exploratory studies, it seems unsatisfactory if the purpose of the simulation study is to investigate relevant conditions motivated by real-world applications. 

We did not encounter an example of DGM modification in our review and assume that this usually occurs without explicit reporting. Reporting problems with particular DGMs and the rationale for changing them in a certain way could be useful for other researchers investigating similar methods, especially if the DGMs originally selected correspond to seemingly realistic scenarios. 

\subsubsection{Method adaptation}
Instead of modifying the DGM, researchers can also modify the methods that show missingness. For example, the tuning parameters of a method can be changed so that the optimization takes longer but is more robust. The simulation study can then be rerun, hopefully producing no more missingness. In our literature review, \citet{Alinaghi2018} did this by switching the random effects variance estimator because the default led to frequent non-convergence.

Another approach is to modify the method only when missingness occurs \citep{Smith2010}. For example, the tuning parameters of a method can be changed iteratively within a simulation repetition until the method converges, as done by \citet{Ojeda2023} who reduced the tuning parameter ``number of knots'' by one in case of non-convergence. 
While the flexibility of the method is reduced in this example, care must be taken when implementing such approaches, as iteratively changing the tuning parameters for one method but not for others could provide an unfair advantage. 
For example, \citet{anderson2021}, who adapted their method to avoid a boundary case in a multistage method, rightfully acknowledged that this procedure ``contaminated'' the method under investigation. Similarly, \citet{Liu2022} replaced one of their methods with a simplified version of the same method whenever an improper solution occurred. They acknowledged that this would also be what they recommend in practice.

\subsubsection{Resource adaptation}
Missingness can arise from insufficient computational resources, for example, because a method takes too long to run (``time-outs'') or runs out of memory (``memory overflows''). One way to handle such missingness is to adapt computational resources. For example, one can adjust the computational environment running the simulation study itself, e.g., by increasing the number of CPU cores physically (by switching to a computer with a more powerful CPU) or remotely (by adjusting the server running the simulation). Alternatively, one can change the computational resource parameters of a method within the simulation code, for instance, by increasing the number of CPU cores over which the method's computations are parallelized.

There are different ways to implement resource adaptation. Ideally, researchers capture errors and rerun individual repetitions with more resources for methods that ran out of resources. 
Allocating sufficient resources to all repetitions (e.g., by increasing CPU cores, RAM, number of iterations, run time) is easier to implement but more expensive, because it also allocates additional resources to repetitions that do not require them. 
In our literature review, \citet{Weber2021},  resolved errors by adjusting the optimizer settings for one method, leading to an absence of missingness on rerun.

\subsubsection{Method replacement}
Several authors have suggested dealing with method missingness by replacing a missing value with a baseline or ``gold standard'' method \citep{Crowther2014, Morris2019, White2024, Wuensch2024}. This approach resembles how, on average, a method would be used in practice -- if a method fails, a data analyst would not give up, but consider an alternative method or a method variant. Method replacement is therefore particularly useful for late-stage simulation studies evaluating the practical utility of specific method usage. However, choosing a replacement is not always straightforward. It also makes interpretation more difficult, since the performance estimates then refer to a mixture of two methods. For example, the empirical standard error of a method then relates to the combination of two methods, which may be quite different from the standard error estimate produced by the method in practice. 
Finally, a risk with this approach is that a method A may appear worse overall than a method B simply because it tends to be missing more often and its performance is worsened by the replacement method in the missing repetitions. Yet A might outperform B in conditions where both converge. To identify this, sensitivity analyses comparing method-/repetition-wise deletion and method replacement approaches are necessary (see Section~\ref{sec:recommendations}).

In our literature review, \citet{Ojeda2023} used such an approach. As mentioned before, they reduced the ``knots'' tuning parameter of a method by one in case of non-convergence. However, if the method still did not converge for three knots, they used logistic regression as this is a common baseline method.

\subsubsection{Imputation}
\label{sec:imputation}
Thinking of missingness in simulation studies as a ``missing data problem'' may suggest using imputation methods, for example, mean imputation or multiple imputation using chained equations \citep[MICE, see e.g.,][]{Rubin1976,vanBuuren2018,Carpenter2023}. 
However, an important difference from missing data in ordinary data analysis is that missing values in simulation studies may not always have an underlying ``true'' value masked by a missingness process. In such a case, using ordinary imputation methods does not make conceptual sense as there is no unobserved value to impute. For example, if a method cannot in theory converge for a given data set (e.g., because of the lack of unique maximum likelihood estimate), then there is no true underlying value that is missing. As discussed earlier, whether or not this is the case is often a matter of definition and depends on whether one is studying a particular implementation or the theoretical concept of a method. An important exception is missingness due to memory overflows and time-outs, which can often be regarded as ``true missingness'' because there is a true value that is unobserved due to a lack of resources to compute it \citep{Wuensch2024}.

Even if a true but missing value exists, there are still important differences between simulation studies and typical missing data problems. For instance, we may not always need an unbiased performance estimate. We may be satisfied with results that are ``not over-optimistically biased'' or ``biased but preserving the rank order of the compared methods''. In that case, one could consider imputation approaches that would not be used in ordinary data analysis such as imputing the worst-case performance of the non-missing methods in the same condition. This approach is easier to communicate than, e.g., a full-blown MICE application, and ``penalizes'' method non-convergence. At the same time, the resulting performance estimates are difficult to interpret, and conclusions may depend on the way the imputation is performed \citep[see][for a demonstration in benchmarking studies]{Niessl2024}.

In our literature review, \citet{Cairns2021} chose to impute the full parameter space as a confidence interval (i.e., a ``worst-case confidence interval'') when an undefined confidence interval was obtained in their simulation study. This approach allows for the analysis of every repetition and might resemble real-world practice, as the worst-case confidence interval represents maximum uncertainty. At the same time, it complicates the interpretation of performance because it increases coverage (as it includes every parameter value) but also increases confidence interval width (as it is the widest possible). Hence, depending on which performance measure is the focus, the approach may be considered liberal (coverage) or conservative (confidence interval width). Apart from this example, imputation-like approaches were rarely used in our literature review, likely because they often make no conceptual sense (because no true value exists), are highly non-standard in simulation studies, and their assumptions (e.g., missing at random) would be difficult to justify to peer reviewers.

\subsubsection{Alternative performance measure}
Even if all methods produce valid outputs, it may still be impossible to estimate method performance from these outputs (performance missingness). For example, model predictions may be so extreme that the empirical mean squared prediction error explodes numerically. 
To still be able to assess performance, one can switch to an alternative performance measure that does not suffer from performance missingness. For instance, one may look at ranks or ``robust'' versions of means and standard deviations. While this approach allows performance to be evaluated, the alternative measure may not be as meaningful as the originally intended one. Using performance measures that trim estimates (e.g., a trimmed mean) or are less sensitive to outliers (e.g., a median) could obscure issues with certain methods by disregarding extreme variability. There is also a risk that neutral comparison of methods may be compromised by selective reporting or cherry-picking, as the choice and implementation of the alternative performance measure may be influenced by the observed results. 

To give an example from our literature review, \citet{Lai2022} tried to avoid the influence of extreme outliers by computing robust versions of bias and empirical standard error based on the trimmed mean and mean absolute deviation, respectively. This required a choice of the trimming proportion, which the authors chose to be 20\%. While the authors noted that they considered this to be a good compromise between robustness and sensitivity to outliers, other choices could also be justifiable. 

\section{Case study}
\label{sec:case-studies}
We illustrate the impact of different approaches to handling missingness using the simulation study by \citet{Carter2019}, which compared seven methods for adjusting for publication bias in meta-analysis. The simulation study employed a fully factorial design of effect size (4) $\times$ between-study heterogeneity (3) $\times$ number of studies (4) $\times$ publication bias (3) $\times$ and questionable research practices (QRPs) (3) factors, resulting in 432 unique conditions with 1000 simulation repetitions in each. The methods were compared on type I error rate, power, root mean squared error, bias, and confidence interval coverage performance measures. However, the study was substantially complicated by four methods (\textit{p}-curve, \textit{p}-uniform, trim-and-fill, and the 3 parameter selection model) showing high rates of non-convergence in certain conditions. For example, for one condition,
the \textit{p}-curve/uniform methods failed to converge 77\% of the time, making interpretation of the performance particularly difficult.

In our literature review, the simulation study would rank among the best-reported studies; \citet{Carter2019} provided per method-condition non-convergence rates, clearly stated that the performance measures were based on method-wise deletion, acknowledged method limitations due to non-convergence, and openly shared code and data for reproducing their results (\url{https://osf.io/rf3ys/}). In the following, we will use these data to reanalyze their simulation study. 

\subsection{Summarizing and understanding missingness}
\begin{figure}[!phtb]
    \centering
    \includegraphics[width=\linewidth]{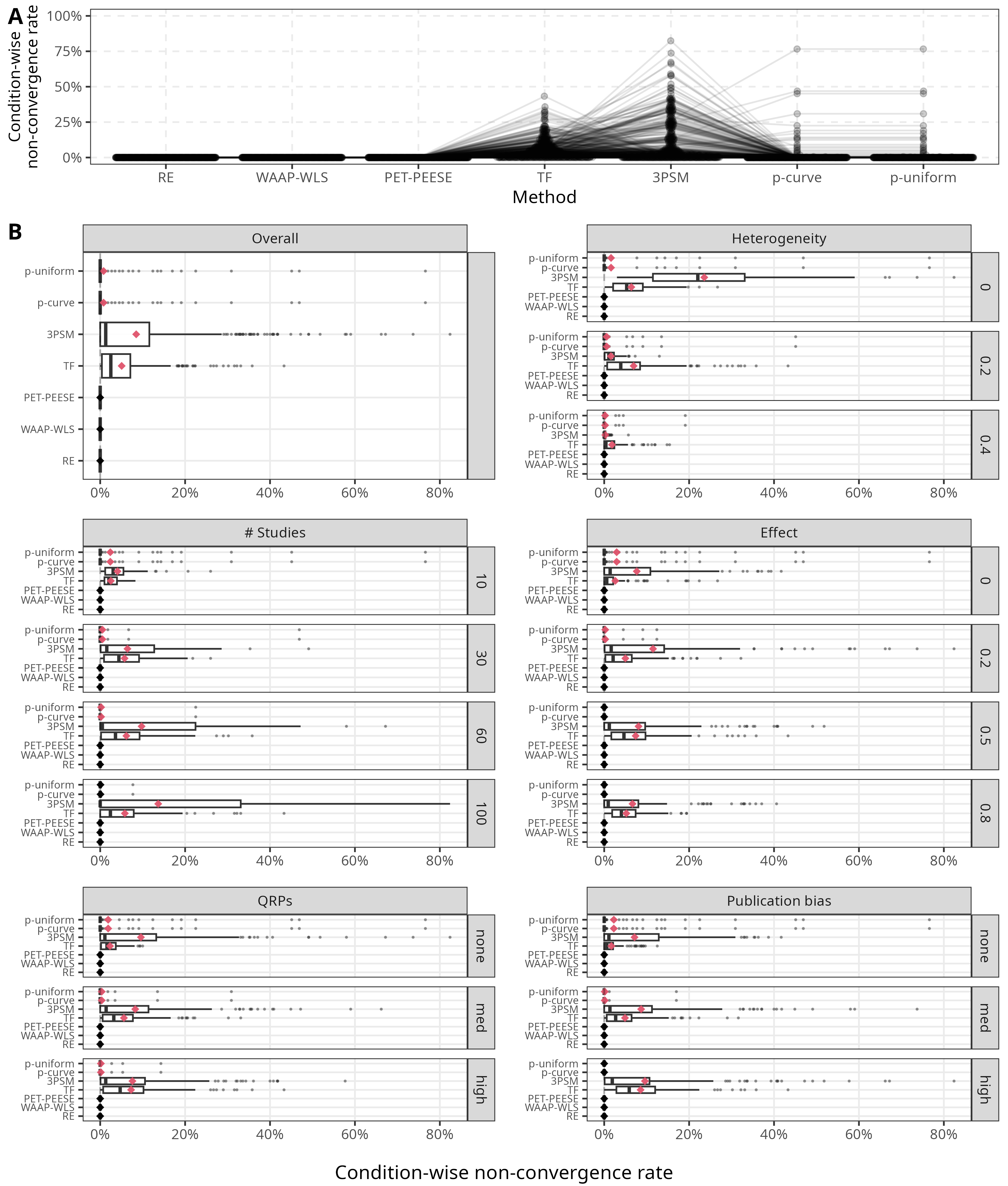}
    \caption{Diagnostic plots for exploring non-convergence in the simulation study from \citet{Carter2019}. Plot A shows a beeswarm plot of the condition-wise non-convergence rates by method with lines connecting conditions. Plot B shows boxplots of the marginal distributions of non-convergence rates stratified by factor and method. The diamond represents the mean non-convergence rate and is colored red if it is greater than 0.1\%.}
    \label{fig:carter2019-exploring}
\end{figure}

Comparing the convergence rates of seven methods across five factors in 432 conditions -- for example, with the four-page Table~2 in the supplementary material of \citet[\url{https://osf.io/vmsxh}]{Carter2019} -- is challenging. We hence explore alternative ways of summarizing and understanding missingness patterns. Figure~\ref{fig:carter2019-exploring}A shows condition-wise non-convergence rates by method with the lines connecting the conditions across methods. We can see that the trim-and-fill (TF), 3 parameter selection model (3PSM), and the \textit{p}-curve/uniform show non-convergence in some conditions, while random effects meta-analysis (RE), weighted average of adequately powered studies - weighted least squares (WAAP-WLS), and precision effect test with standard errors squared (PET-PEESE) always converge. Interestingly, non-convergence rates seem to always coincide between \textit{p}-curve and \textit{p}-uniform (which is evident from parallel lines connecting the two methods). 3PSM and TF, however, show different rates in the same conditions (crossing lines connecting the two methods). Overall, Figure~\ref{fig:carter2019-exploring}A suggests different missingness patterns among the methods which are worth exploring further.

To better understand the influence of the simulation factors on the occurrence of non-convergence, we can visualize the distribution of non-convergence rates per method and simulation factor marginalized over the remaining simulation factors. 
The top-left ``Overall'' panel of Figure~\ref{fig:carter2019-exploring}B is an alternative representation of panel A featuring red diamonds for the mean non-convergence rate and a box plot depicting the distribution of the non-convergence rates. The other panels show the method's marginal non-convergence rates stratified per simulation factor. They provide several insights into non-convergence patterns: (1) while non-convergence rates decrease with the number of studies for \textit{p}-curve/uniform, they increase for TF and 3PSM, (2) a higher degree of QRPs results in less non-convergence for all methods but TF, (3) larger heterogeneity generally leads to lower non-convergence across all methods but TF, which peaks at 0.2 heterogeneity, (4) non-convergence with respect to effect size seems to behave non-monotonically for 3PSM and TF, peaking at 0.2 and 0.5 effect sizes, respectively, but decreasing with increasing effect size for \textit{p}-curve/uniform, and (5) while non-convergence decreases with stronger publication bias for \textit{p}-curve/uniform, it increases for TF and 3PSM.

Domain knowledge allows us to explain some of the patterns of non-convergence. 
The \textit{p}-curve/uniform methods are estimated using only the statistically significant study effect estimates. If too few statistically significant estimates are simulated, \textit{p}-curve/uniform cannot be estimated (i.e., do not converge). This tends to occur more often in conditions without QRPs, publication bias, and an effect of zero. Consequently, \textit{p}-curve/uniform tend to show higher non-convergence rates in these conditions. 
The non-convergence patterns of TF and 3PSM, on the other hand, are more difficult to explain. For example, one would expect non-convergence of 3PSM to decrease with an increasing number of studies, as this typically stabilizes the estimation of the publication selection function. However, an opposite trend is visible, e.g., non-convergence is highest in the condition with 100 studies, an effect of 0.2, heterogeneity of 0, no QRPs, and high publication bias. A closer look at the simulated data sets from this condition could hence be a starting point for further investigations. For example, one could look at interactions of factors that seem to be most associated with non-convergence in a marginal sense, e.g., heterogeneity and the number of studies (figures not shown here but provided in the code repository).

\subsection{Comparing missingness handling approaches}
\begin{figure}[!htb]
    \centering
    \includegraphics[width=0.95\linewidth]{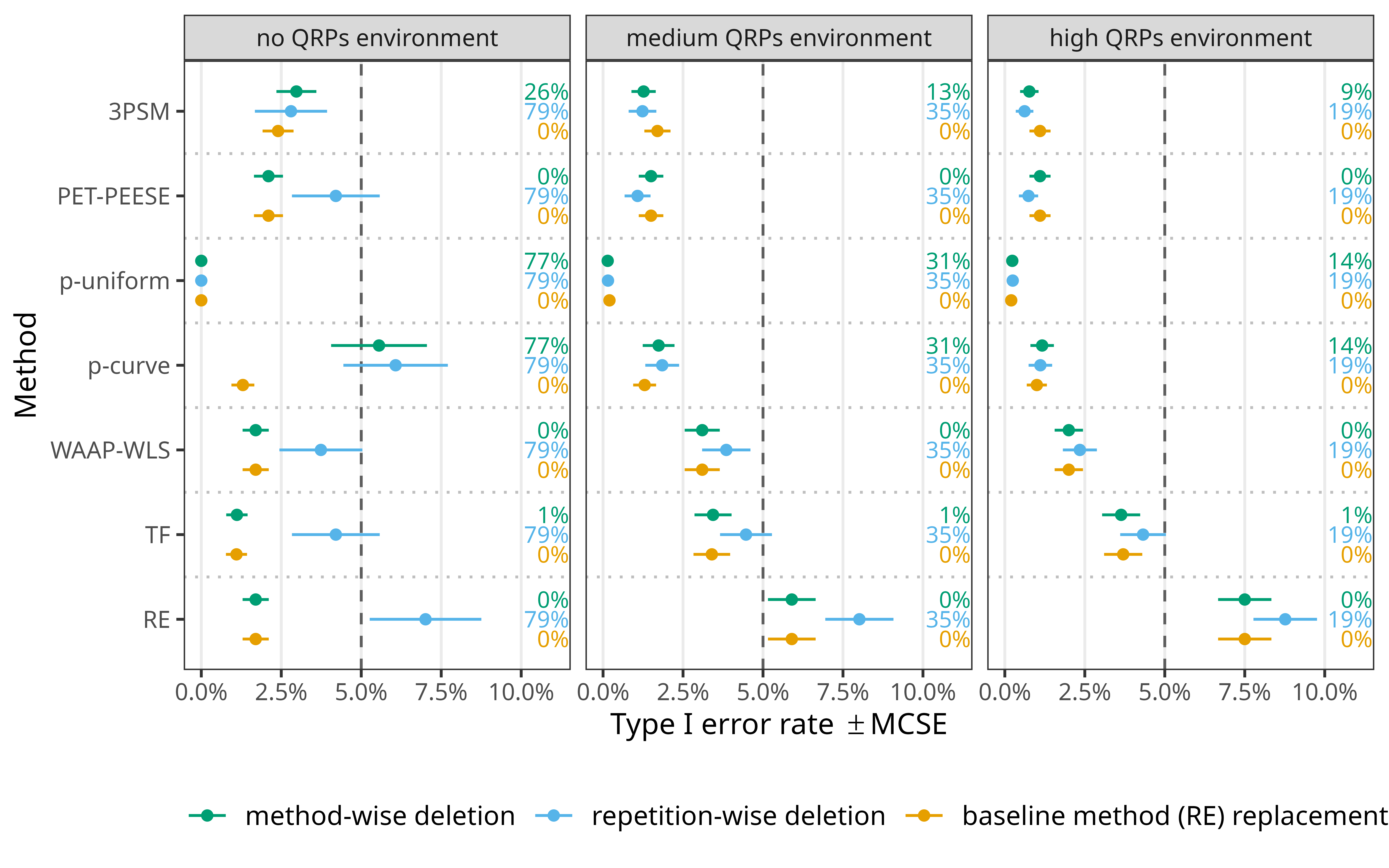}
    \caption{Subset of results from simulation study comparing methods for adjusting for publication bias in meta-analysis by \citet{Carter2019}. Shown are empirical type I error rates with Monte Carlo Standard Errors (MCSEs) along with corresponding non-convergence rates based on different missingness handling approaches (right side) for conditions with no publication bias, no heterogeneity, and 10 studies per meta-analysis.}
    \label{fig:carter2019}
\end{figure}

We will now illustrate how different missingness handling approaches impact the results of the study. For ease of exposition, we focus on the type I error rate performance measure and a trio of conditions varying in the degree of QRPs (none, medium, high) while fixing the remaining conditions (zero effect size, no between-study heterogeneity, 10 studies, and no publication bias). Figure~\ref{fig:carter2019} visualizes the empirical type I error rates (with Monte Carlo Standard Errors (MCSEs), \textit{x}-axis) of the seven methods (\textit{y}-axis) in the three QRPs conditions (vertical panels). The type I error rate of each method is computed under different approaches to handling non-convergence (colors) and is accompanied by the corresponding percentage of missing repetitions (right-hand side). In the following, we will illustrate how these results change when different missingness handling approaches are considered.

First, we consider the results under method-wise deletion (green) as reported by \citet{Carter2019}. We see that all but the RE and \textit{p}-curve methods show type I error rates below the nominal 5\% across the three QRPs conditions. RE shows a slightly too high type I error rate in medium and high QRPs conditions, while the type I error rate of \textit{p}-curve is slightly above 5\% in the no QRPs condition (although 5\% is included within the MCSE error bar). However, the interpretation of all these numbers is conditional on convergence, which drastically differs among the methods. As such, the comparison is not based on the same underlying data and might be unfairly biased because one method has more ``difficult'' or ``easy'' data sets to deal with than another.

Second, in an effort to base method comparison on the same underlying data, one may omit all repetitions with at least one method showing non-convergence (repetition-wise deletion). From the blue statistics in Figure~\ref{fig:carter2019}, we see that such an approach alters the results to some extent: First, it leads to larger MCSEs because more data are discarded than under the method-wise deletion approach. For instance, in the no QRPs condition (left panel), 79\% of the repetitions are discarded, as only in 21\% of all repetitions all methods converged. Furthermore, the estimated type I error rates of the methods can change drastically compared to method-wise deletion. For instance, in the no QRPs condition (left panel), the type I error rate of the RE method changes from about 2\% (well below the nominal value of 5\%) under method-wise deletion to about 7\% under repetition-wise deletion. The choice between method-wise and repetition-wise deletion can thus determine whether a method is judged too conservative (below 5\%) or too liberal (above 5\%). While this comparison ensures that the comparison of method performance is based on the same underlying data, the interpretation is delicate (perhaps even more so than with method-wise deletion) because the estimated type I error rates are conditional on the simultaneous convergence of all compared methods and as such may not be particularly relevant in practice.

Third, to mimic how a non-convergent method would actually be used in practice, we may also replace missing outputs from non-convergent methods with those from a ``baseline'' method. For example, we may use RE meta-analysis as a baseline method since it does not provide any adjustment for publication bias and always converges. Consequently, the approach leads to all methods having no missingness and thus the smallest MCSEs (see the orange statistics in Figure~\ref{fig:carter2019}). At the same time, the interpretation becomes more difficult as the methods are no longer ``pure'', e.g., the \textit{p}-curve/uniform methods in the no QRPs condition (left panel) consist to 77\% of the RE meta-analysis and only 23\% of \textit{p}-curve/uniform. As a result, the estimated type I error rate becomes closer to the type I error rate of RE in conditions with high missingness. For example, the \textit{p}-curve method's estimated type I error rate changes from around 6\% under method-wise deletion to 2\% under RE method replacement. While this approach keeps all repetitions and may be the most relevant for using a method in practice, it obscures the ``true'' method performance by replacing the method when non-convergence occurs.

\subsection{Conclusions}
This case study illustrated common issues and suggested ways to approach missingness in simulation studies. First, visualizing the distribution of the non-convergence rates allowed us to better understand the patterns of non-convergence. We found that non-convergence was systematically affected by simulation factors, with the direction and magnitude varying between methods. 

Second, domain knowledge allowed us to reason about the causes and patterns of non-convergence in some of the methods. For example, the \textit{p}-curve/uniform methods require studies with statistically significant effect estimates as inputs to converge. Consequently, the non-convergence rate of this method was highest in conditions where the true effect was zero and no publication bias or QRPs were present, because more simulated studies then tended to have statistically non-significant \textit{p}-values. 

Third, the comparison of the different handling approaches showed that for the investigated conditions, conclusions and interpretations were mostly consistent between method-/repetition-wise deletion and method replacement, but could differ drastically for certain methods and conditions based on the chosen handling approach. For example, the type I error rate of random effects meta-analysis changed from too conservative to too liberal in some conditions when switching from method-wise deletion or method replacement to repetition-wise deletion. 
Again, domain knowledge can be used to put these results into context. The type I error rate of random effects meta-analysis changed so radically because repetition-wise deletion removes all repetitions with only non-significant \textit{p}-values, thus making the implicit DGM more difficult in the sense that more significant studies are observed even though the true effect is actually zero, and consequently worsening the performance of the method. With this in mind and given the goal of the study (to compare the performance under realistic conditions to give recommendations for application), repetition-wise deletion does not seem to be an appropriate handling approach. Baseline method replacement, on the other hand, seems to be a viable alternative to method-wise deletion chosen by the original authors, as it mimics how an applied researcher might use a publication bias adjustment method -- if the publication bias method does not converge, revert to the unadjusted random effects meta-analysis baseline. 

\section{Recommendations}
\label{sec:recommendations}
We now discuss practical recommendations to help researchers approach missingness in simulation studies; see Table~\ref{tab:recommendations} for a summary.

\begingroup
\renewcommand{\arraystretch}{1.5} 
\begin{table}[!htb]
    \centering
    \caption{Recommendations for approaching missingness in simulation studies.}
    \label{tab:recommendations}
    \begin{tabular}{c p{0.9\textwidth}}
    \toprule 

    \multicolumn{2}{l}{\textit{Design}} \\

    0. & When observing missingness, reconsider methodology and implementation of method and simulation study \\

    \rowcolor{lightgray} 1. & If relevant for method use in practice, conduct simulation studies with missingness as the primary performance measure rather than treating it as a nuisance \\


    2. & Implement simulation code to detect and handle missingness (e.g., capture errors and warnings with exception handling such as \texttt{tryCatch()} in R, save intermediate results, use seeds and repetition-condition identifiers to facilitate debugging, use user-friendly software such as the SimDesign R package \citep{Chalmers2020} to organize and run simulation studies safely and efficiently) \\ 

    \multicolumn{2}{l}{\textit{Analysis}} \\

    \rowcolor{lightgray} 3. & Investigate and discuss potential mechanisms behind missingness (e.g., using visualizations or meta-models) \\

    4. & Align the missingness handling approach with goals of the study (e.g., in more applied simulation studies, consider replacement with baseline method to emulate actual method usage) \\ 
    
    \rowcolor{lightgray} 5. & Perform sensitivity analyses to see how the results change when alternative missingness handling approaches are used (e.g., compare results of method-wise and repetition-wise deletion if both are justifiable)  \\

    \multicolumn{2}{l}{\textit{Reporting}} \\
        
    6. & Quantify and report missingness (ideally by method/condition, use supplement if space is limited) \\
    
    \rowcolor{lightgray} 7. & If no missingness occurred, report this explicitly (e.g., ``\textit{The specified DGM always produced valid data sets, all methods always converged without errors, method performance could always be estimated.}'') \\
    
    8. & Report how missingness was handled and why that approach was chosen (potentially pre-specify the approach before conducting the simulation study to increase neutrality) \\
    
     \rowcolor{lightgray} 9. & Interpret simulation study results in light of missingness and discuss implications for methods' performance and usefulness \\

    10. & Share simulation study code and unaggregated outcome data to disclose technical implementation of missingness handling and to enable secondary analyses \\
     
    \bottomrule
    \end{tabular}
\end{table}
\endgroup

When implementing a simulation study, it is common for initial attempts to produce issues that result in some form of missingness. In this case, the first step should be to rethink the implementation of the DGM, methods, and performance measures. In particular, when a new method is developed and missingness is encountered, this is often indicative of its limitations. Researchers should then reconsider the methodology and implementation of the method to avoid missingness before even attempting to address it in any of the ways discussed in this paper (therefore indicated as ``recommendation 0'' in Table~\ref{tab:recommendations}). However, while iteratively adjusting a method or simulation study to understand and avoid missingness is a normal part of early method development, researchers should also be careful not to selectively choose favorable settings to present an unreliable method as better than it actually is \citep{Pawel2024, Niessl2021}. 

If missingness is a particularly serious problem for certain methods or applications, researchers may consider conducting simulation studies with the primary goal of evaluating missingness rather than treating it as a nuisance. For example, in our literature review, the simulation study by \citet{Cooperman2022} focused entirely on a better understanding of the occurrence of improper solutions (``Heywood'' cases: negative variances or absolute correlations greater than one) in exploratory factor analysis. Such studies can guide data analysts and inform future simulation studies and methodological research.

Following good coding practices when writing simulation study code is critical to detecting and handling missingness \citep{Sigal2016, Morris2019, Chalmers2020, White2024, Siepe2024, Williams2024}. This includes modular design (e.g., separate functions for data generation, method application, and performance assessment), capturing errors and warnings (e.g., using \texttt{tryCatch()} in R when applying a method that may not converge), using seeds for reproducibility, and saving intermediate results for debugging. Dedicated simulation study software, such as the SimDesign R package \citep{Chalmers2020}, can help to safely organize and run simulation studies, including convenience features such as catching errors and parallelization. 

When analyzing the results from their simulation study, it is crucial that researchers always quantify, report, and investigate missingness. This enables readers to assess whether or not missingness poses a problem for the interpretation of the results. Even if no missingness occurred, it is recommended to report this explicitly. Visualizations 
or meta-models can help in reporting and improve understanding of missingness patterns. In addition, it is often advisable to compare the properties of the simulated data sets in missing and non-missing repetitions within a condition to understand potential properties of the data that might lead to missingness. 

If missingness occurs, a missingness handling approach is required to assess performance. It is important to report how missingness was handled and provide justifications for why that approach was chosen. A more neutral evaluation can be achieved by pre-specifying the missingness handling approach before conducting the simulation study \citep{Siepe2024}. The missingness handling approach should be aligned with the goals of the simulation study. For example, if the study aims to compare the performance of methods as employed in practice, one might use a method replacement approach (i.e., missing outputs are replaced with a baseline method), since this emulates the actual method use. In contrast, method replacement may not be advisable in simulation studies aimed at better understanding the theoretical properties of a method (e.g., the asymptotic bias of an estimator), because the resulting performance estimate does not refer to a single method anymore but a mixture of two methods. While simulation studies conducted in ``early-stage'' methodological research tend to focus more on methods as a theoretical concept and late-stage methodological research tends to focus more on specific implementations of methods, the distinction can sometimes be quite subtle \citep{Heinze2024}.  

To explore the potential impact of the choice of the missingness handling approach, one may also perform sensitivity analyses to see how results change when alternative approaches are used (e.g., repetition-wise instead of method-wise deletion if both are justifiable). Such analyses become more important with increasing missingness. However, it is difficult to judge ``how much missingness'' is required until careful handling is necessary. Sensitivity analyses can also be useful for providing more differentiated method recommendations. For example, two result tables could be presented: one showing performance estimates based on method-wise deletion and another showing estimates based on gold standard replacement. This could then inform two-part recommendations, such as using a certain method when convergence occurs (when it performs best) but using another method when convergence does not occur.

It is important to consider missingness when interpreting method performance and making recommendations. For example, even if a method shows excellent performance in the repetitions where it converges, the method may not be useful in practice if it often fails to converge under relevant conditions. Researchers should then not recommend the method without stating this caveat, and should provide recommendations on what to do when the method fails to converge (for example, done by \citet{Johal2023}, part of our literature review).

Finally, code sharing is another important step in improving the reproducibility of simulation studies. It also allows other researchers to examine the technical implementation of the missingness handling and to perform alternative analyses, e.g., using a different handling approach. To save other researchers from having to re-run the entire study (which can often take days or weeks), it is also advisable to share data in some intermediate form (e.g., parameter estimates or other method outputs for each repetition of the simulation study, or files with all error messages as in \citealp{Weber2021}). 

\section{Discussion}
\label{sec:discussion}

Our literature review demonstrated that issues related to missingness are rarely reported or discussed in simulation studies published in prominent methodological journals. This contradicts our personal experience and discussions with colleagues suggesting that these issues are common. The way missingness is handled can have substantial consequences for the analysis of simulation study results, and blur interpretations and conclusions. We therefore believe that missingness deserves more attention and should become a more routine consideration in the design, analysis, and reporting of simulation studies. We have provided detailed recommendations on how this could be implemented in practice. As a bare minimum, we recommend that researchers quantify and report missingness and how it was handled, and provide code to reproduce the simulation study and its analysis. This is in line with recent calls for higher standards of reproducibility and replicability of simulation studies \citep{Boulesteix2020, Lohmann2022, Wrobel2024, Luijken2024, Williams2024}.

In our literature review, it is possible that certain DGMs and methods under investigation were generally not prone to missingness. In such cases, acknowledgements such as ``all methods converged under all conditions'' might have seemed too obvious to the authors. However, such a judgment requires an in-depth analysis of each study and its methods, which was infeasible for the review at hand. Simulation studies investigating the performance of such ``never-fail'' methods and DGMs may still suffer from missingness due to coding errors or other implementation problems. We believe therefore that, even in these situations, it may be useful to the reader to report transparently that no missingness was observed.

Exploring and handling of missingness in simulation studies is not straightforward. Future research could extend the work of \citet{Gasparini2021} and explore different types of diagnostics and visualizations that may facilitate this process, and provide implementations that are interoperable with established simulation study software (e.g., the SimDesign R package from \citealp{Chalmers2020} or the simsum Stata module from \citealp{White2010}). 

Several of our recommendations relate to code design and other computational aspects. We recognize that implementing these suggestions will require computational expertise on the part of the researcher. This needs to be addressed through better training, the availability of more tutorials for different programming environments \citep[such as the R Tutorial in][]{Sigal2016}, and support staff at the institutional level.

As most of our recommendations relate to transparency, reporting, and exploration, these expectations could be translated into author guidelines, peer review manuals, and journal requirements. For example, journals with a majority of articles containing simulation studies (e.g., JASA or SiM) might consider providing simulation reporting guidelines, similar to their reproducibility guidelines \citep{Wrobel2024}. As our case study illustrated, many missingness handling approaches can be implemented ``post-hoc'' and do not require re-running all code. Reviewers should thus not shy away from requesting them to ensure an accurate interpretation of results. Similarly, journals should consider making code and data sharing mandatory. This would also facilitate further meta-research on missingness or other aspects of simulation studies.

Despite the overarching principles and recommendations we have outlined, missingness handling will often need to be done on a case-by-case basis. Nevertheless, a consensus on default reporting and handling approaches may be possible in some methodological subfields. We hope that our article will raise awareness of the issue and stimulate such discussions.

\section*{Software and data}
\label{sec:code}
Data analysis was carried out using R \citep[version 4.4.1,][]{R2024}. We used the tidyverse packages \citep{tidyverse2019} for data wrangling and ggplot2 \citep{Wickham2016} for visualization. Data and code to reproduce our analyses, information on the computational environment, and additional results from our literature review are available in the online supplement (\anonymize{\url{https://github.com/SamCH93/missSim}}). A snapshot of the repository at the time of writing is archived on Zenodo \anonymize{(\url{https://doi.org/10.5281/zenodo.13846651})}. The simulation summary data from \citet{Carter2019} were downloaded from \url{https://github.com/nicebread/meta-showdown}.

\section*{Author contributions}
The authors made the following contributions: \anonymize{SP: Conceptualization, Methodology, Formal Analysis, Software, Investigation, Data Curation, Visualization, Writing -- original draft, Writing -- review \& editing; BSS: Conceptualization, Methodology, Formal Analysis, Software, Investigation, Data Curation, Visualization, Writing -- original draft, Writing -- review \& editing; FB: Conceptualization, Methodology, Formal Analysis, Software, Investigation, Data Curation, Visualization, Writing -- original draft, Writing -- review \& editing; AL: Methodology, Formal Analysis, Software, Investigation, Data Curation, Visualization, Writing -- original draft, Writing -- review \& editing;}

\section*{Acknowledgments}
We thank \citet{Carter2019} for openly sharing their code and data. We thank \anonymize{Rolf Groenwold} for helpful comments on a draft of the manuscript. 
We thank the associate editor and two anonymous referees for constructive and helpful comments that greatly improved the manuscript. 
Our acknowledgment of these individuals does not imply their endorsement of this article.

\section*{Disclosure Statement}
The authors report that there are no competing interests to declare.

\bibliographystyle{icml2024}
\bibliography{bibliography}

\section*{Appendix: Inter-rater agreement}
\label{sec:appendix}
\begin{figure}[!htb]
    \centering
    \includegraphics[width=\linewidth]{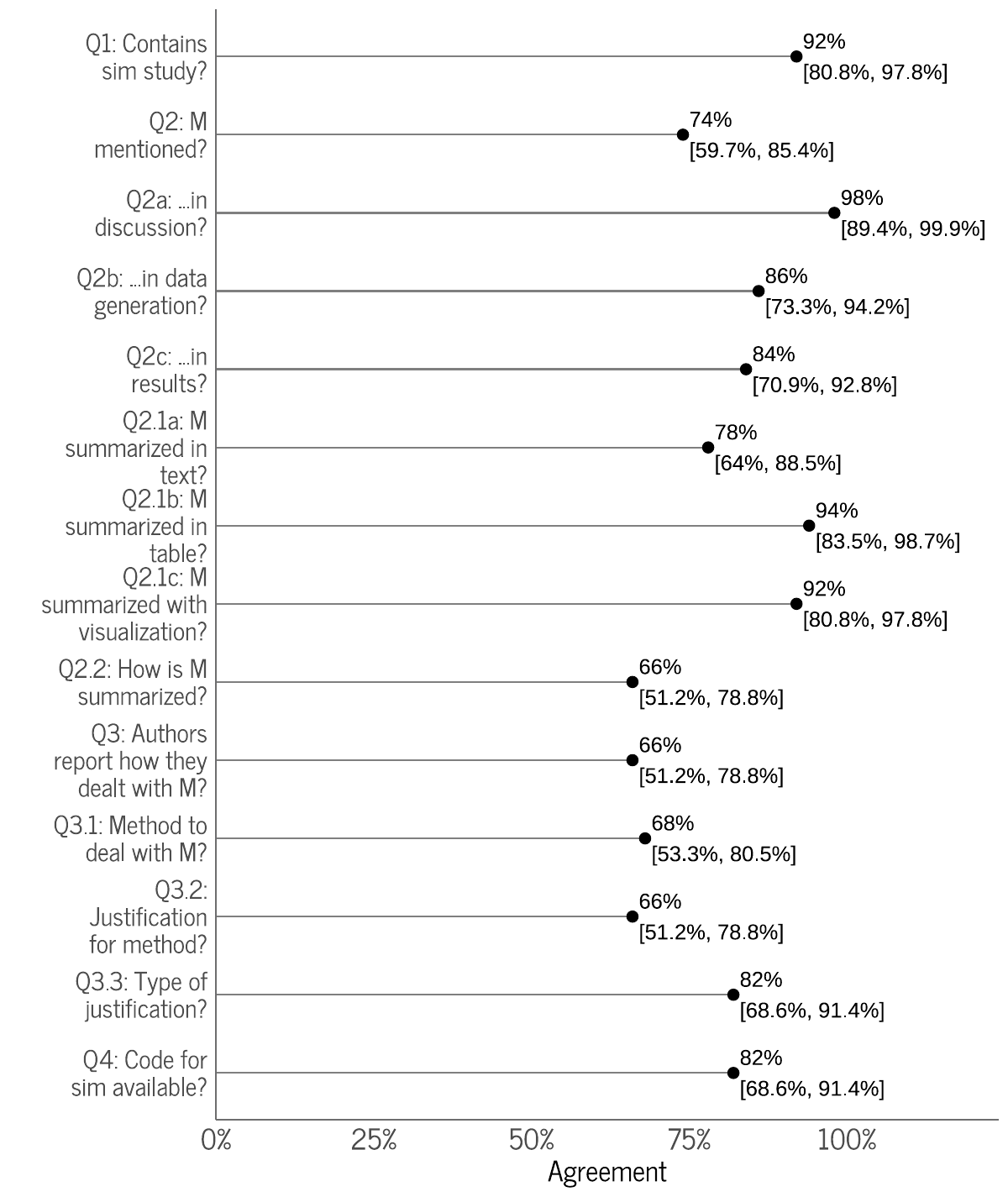}
    \caption{Agreement percentage for 50 studies with ``low'' or ``medium'' confidence. ``M'' stands for ``Missingness''. Intervals in brackets denote 95\% confidence intervals for proportions as calculated with the Clopper-Pearson method as implemented in \texttt{stats::binom.test} in R \citep{R2024}.}
    \label{fig:agreement-results}
\end{figure}

We provide a summary of the inter-rater agreement for a subset of studies in Figure~\ref{fig:agreement-results}. Overall, the agreement seems to be acceptable to good. Since we used the studies that we found most difficult to code for our assessment of agreement, these proportions can be interpreted as a lower bound on agreement. The lowest levels of agreement were found for questions on missingness summarization and dealing with missingness. The unclear handling of missingness was often the reason why we found these studies difficult to code. For example, in one of the studies where we disagreed on parts of question 3, several causes and strategies for handling non-convergence were hinted at, but not described in detail. 

\end{document}

%% file: missingness-classification-tikz.tex
\begin{tikzpicture}[thick,every node/.style={scale=0.95}]

\node [rectangle, draw, rounded corners = 0.5em] (truth)
{\begin{tabular}{c} 
  \textbf{Underlying} \\
  \textbf{Truth}
\end{tabular}};

\node [rectangle, draw, rounded corners = 0.5em] (simdat) [right = 9em of truth]
{\begin{tabular}{c} 
  \textbf{Simulated} \\
  \textbf{Data}
\end{tabular}};

\node [rectangle, draw, rounded corners = 0.5em] (output) [right = 9em of simdat]
{\begin{tabular}{c} 
  \textbf{Analysis} \\
  \textbf{Output}
\end{tabular}};

\draw [->] (truth) -- node [above] (dgp)
{\begin{tabular}{c} 
  Data-Generating \\ 
  Mechanism 
\end{tabular}}
(simdat);

\draw [->] (simdat) -- node [above] (analysis)
{\begin{tabular}{c} 
  Statistical \\ 
  Analysis 
\end{tabular}}
(output);

\draw [->] (output.south) to [bend left = 15] node [below, sloped] (performance)
{\begin{tabular}{c}
  Performance \\ 
  Evaluation 
\end{tabular}}
(truth.south);

\node [rectangle, draw, rounded corners = 0em, fill = gray!10] (methods) [above = 1em of analysis]
{\footnotesize 
\begin{tabular}{l}
  \multicolumn{1}{c}{\textbf{Method Missingness}} \\
  Method does not produce valid output \\
\end{tabular}
};

\node [rectangle, draw, rounded corners = 0em, fill = gray!10] (params) [above = 1em of dgp]
{\footnotesize 
\begin{tabular}{l} 
  \multicolumn{1}{c}{\textbf{DGM Missingness}}\\
  Simulated data set is not valid \\ 
\end{tabular}
};

\node [rectangle, draw, rounded corners = 0em, fill = gray!10] (metrics) [right = 4em of performance]
{\footnotesize 
\begin{tabular}{l} 
  \multicolumn{1}{c}{\textbf{Performance Missingness}} \\
  Metric cannot be computed \\ 
\end{tabular}
};

\draw [->] (params.south) to (dgp);
\draw [->] (methods.south) to (analysis);
\draw [->] (metrics.west) to (performance.east);

\end{tikzpicture}